\documentclass[conference]{IEEEtran}
\pdfoutput=1

\usepackage{latexsym}
\usepackage{amsmath}
\usepackage{amssymb}
\usepackage{amsthm}
\usepackage{graphicx}
\usepackage{caption}
\usepackage{subcaption}
\usepackage{accents}
\usepackage[inline]{enumitem}
\usepackage{xcolor}
\usepackage{array}
\usepackage{multirow}
\usepackage{cite}
\usepackage{varwidth}

\usepackage[hyphens]{url}
\usepackage{hyperref}

\usepackage{tikz,pgfplots}
\usetikzlibrary{arrows,positioning,automata,calc}
\usetikzlibrary{plotmarks,patterns}

\pgfplotsset{
    legend image with text/.style={
        legend image code/.code={%
            \node[anchor=center] at (0.3cm,0cm) {#1};
        }
    },
}

 \newcounter{MYtempeqncnt}

\newcommand{\E}{\mathbb{E}}

\def \blue{\color{blue}}
\definecolor{byzantium}{rgb}{0, 0, 0.5}

\def \w{{\fontfamily{cmtt}\selectfont \text{W}}}
\def \m{{\fontfamily{cmtt}\selectfont \text{M}}}

\newtheorem{theorem}{Theorem}

\newtheorem{claim}{Claim}
\newtheorem{example}{Example}

\newtheorem{corollary}{Corollary}

\IEEEoverridecommandlockouts

\begin{document}

\newlength\figureheight
\newlength\figurewidth

\title{Minimizing Latency for Secure Distributed Computing \vspace{-0.2cm}}
\author{
\IEEEauthorblockN{Rawad Bitar, Parimal Parag, and Salim El Rouayheb
\thanks{R. Bitar and S. El Rouayheb are with the ECE department of Illinois Institute of Technology. P. Parag is with the ECE department of the Indian Institute of Science. Emails: rbitar@hawk.iit.edu, parimal@ece.iisc.ernet.in, salim@iit.edu.}}
\vspace{-1.5cm}
}

\maketitle

\begin{abstract}
We consider the setting of a master server who possesses confidential data (genomic, medical data, etc.) and wants to run intensive computations on it, as part of a machine learning algorithm for example. The master wants to distribute these computations to untrusted workers  who have volunteered or are incentivized  to help with this task. However,  the data must be kept  private (in an information theoretic sense) and not revealed to the individual workers. The workers may be busy, or even unresponsive, and will take a random time to  finish the task assigned to them. We are interested in reducing the aggregate delay experienced by the master. We focus on linear computations as an essential operation in many iterative algorithms. A known solution is to use a linear secret sharing scheme to divide the data into secret shares on which the workers can compute. We propose to use instead new secure codes, called Staircase codes, introduced previously by two of the authors. We study the delay induced by  Staircase codes which is always less than that of secret sharing.   The reason is that secret sharing schemes need\ to wait for the responses of a fixed fraction of the workers, whereas Staircase codes offer more flexibility in this respect. For instance, for codes with rate $R=1/2$ Staircase codes can lead to up to $40\%$ reduction in delay compared to secret sharing.
\end{abstract}

\section{Introduction}
We consider the setting of distributed computing in which a server $\m$, referred to as Master,  possesses  confidential  data, such as personal information of online users, genomic and medical data etc., and wants  to perform  intensive computations on it.  $\m$ wants to divide these computations into smaller computational tasks and distribute them to $n$ worker machines that can perform these smaller tasks in parallel. The workers then return their results to the master, who can  process them to obtain the result of its original task. The well celebrated MapReduce \cite{mapreduce} framework falls under this model and is implemented in many computing clusters.

 In this paper, we are interested in applications in which the worker machines do not belong to the same system or cluster as the master. Rather, the workers are online computing machines that can be  hired or can volunteer to help the master in its computations. Existing applications that fall under this model include the SETI@home project for search for extraterrestrial intelligence \cite{setiah}, the folding@home project for disease research that simulates protein folding \cite{fah} and Amazon mechanical turk\footnote{Amazon mechanical turk hires humans to perform tasks. But, one can imagine a similar application where computing machines are hired.} \cite{amt}. The additional constraint that we worry about here, and which does not exist in the previous applications, is that  the workers cannot be trusted with the sensitive data, which must remain hidden from them. Our privacy constraint  is information theoretic, meaning that each worker must obtain zero information about the data irrespective of its computational power. We choose information theoretic privacy instead of homomorphic encryption, due to the high computation and memory overheads of the latter \cite{homomorphic2}.

We focus on  linear computations (matrix multiplication) since they form a basic building block of many iterative algorithms. {The workers introduce random delays due to the difference of their workloads or network congestion. This causes the Master to wait for the slowest workers, referred to as stragglers in the distributed computing community \cite{DB13,AKGSLSH10,speeding}. In addition, some workers may never respond. Our goal is to reduce the delay at the Master caused by the workers.}

Privacy can be achieved by encoding the data using a linear secret sharing codes \cite{AF10} as   illustrated  in Example~\ref{ex:intro}. However, these codes are not specifically designed to minimize latency as we will highlight later.
\begin{example}
\label{ex:intro}
Let the matrix $A$ denote the data set owned by $\m$ and let $\mathbf{x}$ be a given vector. $\m$ wants to compute $A\mathbf{x}$ . Suppose that $\m$ gets the help of $n=3$ workers out of which at most \mbox{$n-k=1$} may be unresponsive. $\m$ generates a random matrix $R$ of same dimensions as $A$ and over the same field and encodes $A$ and $R$ into 3 shares $S_1=R$, $S_2=R+A$ and $S_3=R+2A$ using a secret sharing scheme \cite{S79,McESa81}. First, $\m$ sends  share $S_i$ to worker $\w_i$ (Figure~\ref{fig:introfig1}) and then sends $\mathbf{x}$ to all the workers. Each worker computes $S_i\mathbf{x}$ and sends it back to $\m$ (Figure~\ref{fig:introfig2}). $\m$ can decode $A\mathbf{x}$ after receiving any $k=2$ responses. For instance, if the first two workers respond, $\m$ can obtain $A\mathbf{x}= S_2\mathbf{x}-S_1\mathbf{x}$. No information about $A$ is revealed to the workers, because $A$ is one-time padded by $R$. \vspace{-0.3cm}\begin{figure}[h!]
\centering
\begin{minipage}[t]{0.22\textwidth}
\raggedright
\resizebox{!}{0.75\textwidth}{
\begin{tikzpicture}[>=stealth', auto,
 triangle/.style = {fill=white, regular polygon, regular polygon sides=3 },]
  \definecolor{lightgray}{rgb}{0.83, 0.83, 0.83}
\def\mx{3}
\def\my{-4}
\tikzstyle{server} = [fill=black!10, rectangle, rounded corners=4mm, draw,minimum width=2em, minimum height=2.5em]
\node[inner sep=0] (s1) at (-4.3,0) {\includegraphics[height=9mm]{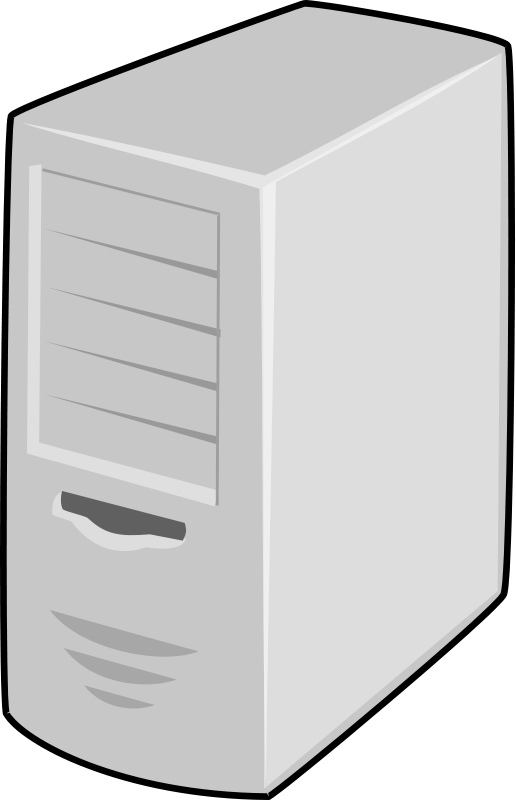}} node[below=0cm of s1,color=byzantium] {Master} node[left=0cm of s1,color=byzantium] {$\m$};

\node[server,right=0.8cm of s1,font=\footnotesize] (bb) {Secret sharing}; 
\node[above=0.8cm of bb, font=\footnotesize] (rm) {Randomness};

\node[inner sep=0pt,right =0.75 of bb,font=\footnotesize] (us2){\includegraphics[height=9mm]{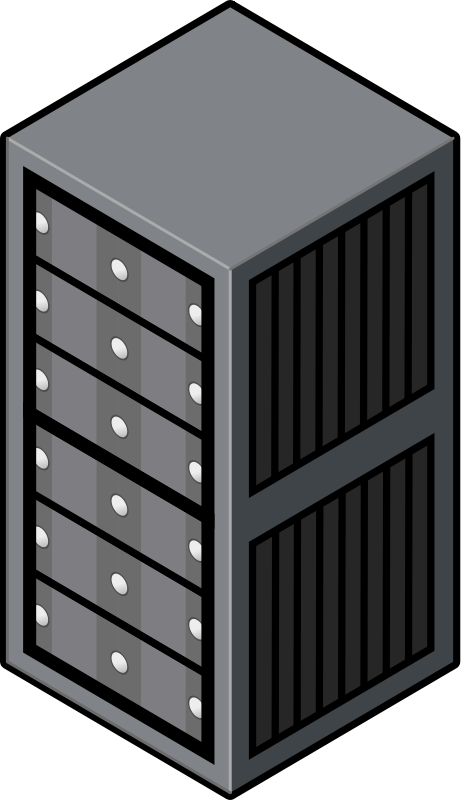}}
node[right =0 of us2,font=\small,color=byzantium] {$\w_2$};

\node[inner sep=0,above=0.35 cm of us2,font=\footnotesize] (us1) {\includegraphics[height=9mm]{spymobile.png}}
node[right= 0 cm of us1,font=\small,color=byzantium] {$\w_1$};

\node[inner sep=0,below= 0.35 cm of us2,font=\footnotesize] (us3) {\includegraphics[height=9mm]{spymobile.png}}
node[right=0 of us3,font=\small,color=byzantium] {$\w_3$} ;
 
\path[every node/.style={},color=black,->]
(s1) edge node[midway,below=0,font=\footnotesize] {$A$} node[midway,above=0,font=\footnotesize] {Data} (bb) 
   (bb.north east)    	edge  node[midway,above=0,sloped,font=\footnotesize] {$S_1$}  (us1) 
   (bb.east)		edge  node[midway,above=0,sloped,font=\footnotesize] {$S_2$}  (us2)
   (bb.south east)	edge  node[midway,below=0,sloped,font=\footnotesize] {$S_3$}  (us3)
   (rm) 			edge  node[midway,right=0,font=\footnotesize] {$R$} (bb);
   
\node[above=0 of us1,color=byzantium] {Workers};
\end{tikzpicture}
}\vspace{-0.4cm}
\captionsetup{subtype,width=0.95\textwidth}
\caption[width=0.85\textwidth]{$\m$ encodes $A$ into $3$ secret shares $S_1,\ S_2,\ S_3$ and sends them to the workers.}
\label{fig:introfig1}
\end{minipage} \hfill\begin{minipage}[t]{0.22\textwidth}
\raggedleft
\resizebox{!}{0.75\textwidth}{
\begin{tikzpicture}
\node[inner sep=0] (s) at (0.7,0) {\includegraphics[height=9mm]{mobile1.png}} node[left=0cm of s,color=byzantium] {$\m$};
\node[inner sep=0pt,right =3.5 of s,font=\footnotesize] (us2){\includegraphics[height=9mm]{spymobile.png}}
node[left =-0.1 of us2,font=\small,color=byzantium] {} node[below =0 of us2,font=\footnotesize] {$S_2$} node[right =0 of us2,font=\small,color=byzantium] {$\w_2$};;

\node[inner sep=0,above= 0.6 cm of us2,font=\footnotesize] (us1) {\includegraphics[height=9mm]{spymobile.png}}
node[left=-0.1cm of us1,font=\small,color=byzantium] {} node[below=0 of us1,font=\footnotesize] {$S_1$} node[right =0 of us1,font=\small,color=byzantium] {$\w_1$};;

\node[inner sep=0,below= 0.6 cm of us2,font=\footnotesize] (us3) {\includegraphics[height=9mm]{spymobile.png}}
node[left=-0.1 of us3,font=\small,color=byzantium] {} node[below= 0cm of us3,font=\footnotesize] {$S_3$} node[right =0 of us3,font=\small,color=byzantium] {$\w_3$};;
\path[every node/.style={},color=black,->]
   (s.north east)    	edge[bend left]  node[midway,above left=0 and -0.4 cm,sloped,font=\footnotesize] {$\mathbf{x}$} (us1) 
   (s.east)		edge[bend left]  node[midway,below=0,sloped,font=\footnotesize] {$\mathbf{x}$}  (us2)
   (s.south east)		edge[bend right]  node[midway,below=0,sloped,font=\footnotesize] {$\mathbf{x}$}   (us3)
			
   (us1)		edge[dashed]  node[midway,above =0, sloped,font=\footnotesize]  {$S_1\mathbf{x}$}  	(s.north east) 
   (us2)		edge[dashed]  node[midway,below =0, sloped,font=\footnotesize]  {$S_2\mathbf{x}$}  	(s)
   (us3)		edge[dashed]  node[midway,below =0, sloped,font=\footnotesize]  {$S_3\mathbf{x}$}  	(s.south east)
   ;

\end{tikzpicture}
}\vspace{-0.4cm}
\captionsetup{subtype,width=0.95\textwidth}
\caption{$\m$ sends $\mathbf{x}$ to the workers. Each worker $\w_i$ computes $S_i\mathbf{x}$ and sends the result to $\m$.}
\label{fig:introfig2}
\end{minipage}
\caption{Secure distributed matrix multiplication with $3$ workers.}\label{fig:intro1}
\end{figure}
\vspace{-0.2cm}
\end{example}

The delay  experienced by $\m$ in the previous example results from the fact it has to wait until  $k=2$ workers finish their whole tasks in order to decode $A\mathbf{x}$, even when the $3$ workers are all responsive.  This is due to the fact that classical secret sharing codes are designed for the worst-case scenario of one worker being unresponsive. We overcome this limitation by using Staircase codes which were introduced in \cite{BR16} and are explained in the next example.

\begin{example}[Staircase code]
\label{ex:intro2}
Consider the same setting as Example~\ref{ex:intro}. Instead of using a classical secret sharing code, $\m$ now encodes $A$ and $R$ using the Staircase code given in Table~\ref{tab:CESS}.
\begin{table}[h!] 
\centering
\begin{tabular}[h!]{c|c|c}
Worker 1 & Worker 2 & Worker 3 \\ \hline
$A_1+A_2+R_1$ & $A_1+2A_2+4R_1$ &$A_1+3A_2+4R_1$\\
\blue $R_1+R_2$ & \blue $R_1+2R_2$ & \blue $R_1+3R_2$ \\
\end{tabular}
 \caption{The shares sent by $\m$ to each worker. All operations are in $GF(5)$.} \label{tab:CESS}
 \vspace{-0.2cm}
\end{table}
The Staircase code requires $\m$ to divide the matrices $A$ and $R$ into $A=\begin{bmatrix}A_1& A_2\end{bmatrix}^T$ and $R=\begin{bmatrix}R_1& R_2\end{bmatrix}^T$. In this setting, $\m$ sends two subshares to each worker, hence each task consists of $2$ subtasks. The master sends $\mathbf{x}$ to all the workers. Each worker  multiplies the subshares by $\mathbf{x}$ (going top to bottom) and sends each multiplication back to $\m$ independently. Now, $\m$ has two possibilities for decoding:
  \begin{enumerate*}[label=\emph{\arabic*)}] \item $\m$ receives the first subtask from all the workers, i.e., receives $(A_1+A_2+R_1)\mathbf{x}$, $(A_1+2A_2+4R_1)\mathbf{x}$ and $(A_1+2A_2+4R_1)\mathbf{x}$ and decodes $A\mathbf{x}$ which is the concatenation of $A_1\mathbf{x}$ and $A_2\mathbf{x}$. Note that $\m$ decodes only $R_1\mathbf{x}$ and does not need to decode $R_2\mathbf{x}$. \item $\m$ receives all the subtasks from any $2$ workers and decodes $A\mathbf{x}$. Here $\m$ has to decode $R_1\mathbf{x}$ and $R_2\mathbf{x}$.
 \end{enumerate*}
One can check that no information about $A$ is revealed to the workers.\end{example}

Under an exponential delay model for each worker, we show that the Staircase code given in Example~\ref{ex:intro2} can lead to a $25\%$ improvement in delay over the secret sharing code given in Example~\ref{ex:intro}. Our goal is to give a general systematic study of the delay incurred by Staircase codes and compare it to classical secret sharing codes.

{\em Related work:} 
Straggler mitigation and privacy concerns are studied separately in the literature.
In \cite{TOFEC} Liang et al.\ adaptively encoded the tasks depending on the workload at the workers' end. Lee et al. \cite{speeding} used MDS codes to mitigate stragglers in {\em linear} distributed machine learning algorithms. Tandon et al. \cite{tandon2016gradient} introduced new codes for straggler mitigation in distributed gradient descent algorithms. Li et al. \cite{li2016fundamental} studied the effect of the workers' computation load on the communication complexity.

On the other hand, privacy concerns have been studied in the machine learning literature, see e.g., \cite{secureml1,secureml2,secureml3}. The main model assumes that several parties owning private data sets want to train a model based on all the data sets without revealing them, e.g., \cite{NWIJBT13,GSBRDZE16}. However, the techniques extensively rely on cryptographic assumptions and secure multi-party computation. Atallah and Frikken \cite{AF10} studied the problem of distributively multiplying two private matrices assuming that $k-1$ workers can collude (with $k^2<n^2$). The provided solution ensures information theoretic privacy, but does not account for straggler mitigation.
Another related problem is federated learning \cite{federated}. A large number of users own different amounts of data and a central server aims to train a high-quality model based on all the data with the smallest communication complexity. However, privacy is ensured by keeping the data local to the users.

{\em Contributions:} In this paper, we consider the model in which $\m$ owns the whole data set on which it wants to perform a distributed linear computation. We introduce a new approach for securely outsourcing the linear computations to $n$ workers which do not own any parts of the data. The data set is to be kept private in an information theoretic sense. 
We assume that at most $n-k,\ k<n$, workers may be unresponsive, the remaining respond at random times. {This is similar to the straggler problem}. We study the master's waiting time, i.e., the aggregate delays caused by the workers, under the exponential model when using Staircase codes. More specifically, we make the following contributions:  \begin{enumerate*}[label=\emph{(\roman*)}] \item we derive an upper bound and a lower bound on the mean waiting time; \item we derive an integral expression leading to the CDF of the waiting time and use this expression to find the exact mean waiting time for the cases when $k=n-1$ and $k=n-2$; and \item we compare our approach to the approach using secret sharing and show that for high rates, $k/n$, and small number of workers our approach saves about $40\%$ of the waiting time.\end{enumerate*} Moreover, we ran simulations to check the tightness of the bounds and show that for low rates our approach saves at least $10\%$ of the waiting time for all values of $n$.

\section{System Model} \label{sec:sysmod}
We consider a server $\m$ which wants to perform intensive computations on confidential data represented by an $m \times \ell$ matrix $A$ (typically $m>>\ell$). $\m$ divides these computations into smaller computational tasks and assigns them to $n$ workers $\w_i$, $i=1,\dots,n$, that can perform these tasks in parallel.

\indent {\em Computations model:} We focus on linear computations. The motivation is that  a building block in several iterative machine learning algorithms, such as gradient descent, is the multiplication of $A$ by a  sequence of  $\ell \times 1$ attribute vectors $\mathbf{x}^1, \mathbf{x}^2, \dots$. In the sequel, we focus on the multiplication $A\mathbf{x}$ with one attribute vector $\mathbf{x}$.

{\em Workers model:} The workers have the following properties: 
\begin{enumerate*}[label=\emph{\arabic*)}]
\item At most $n-k$ workers may be unresponsive. The actual number of unresponsive workers is unknown a priori.
\item {The responsive workers  incur random delays while executing the task assigned to them by $\m$ resulting in what is known as the straggler problem \cite{DB13,AKGSLSH10,speeding}.} We model all the delays incurred by each worker by an independent and identical exponential random variable. \item The workers do not collude, i.e., they do not share with each other the data they receive from $\m$. This has implications on the privacy constraint described later.\end{enumerate*}

{\em General scheme:} $\m$ encodes $A$, using randomness, into $n$ shares $S_i$ sent to worker $\w_i$, $i=1,\dots,n$. Any $k$ or more shares can decode $A$. The workers obtain zero information about $A$, i.e., $H(A|S_i)=H(A)$ for all $i\in\{1,\dots,n\}$.

\noindent At each iteration, the master sends $\mathbf{x}$ to all the workers. Then, each worker computes $S_i\mathbf{x}$ and sends it back to the master. Since the scheme and the computations are linear, the master can decode $A\mathbf{x}$ after receiving enough responses\footnote{In some cases the attribute vectors $\mathbf{x}^j$ contain information about $A$, and therefore need to be hidden from the workers. We describe in \cite{BPR17} how our scheme can be generalized to such cases.}. We refer to such scheme as an $(n,k)$ system.

{\em Encoding:} We consider classical secret sharing codes \cite{S79,McESa81} and universal Staircase codes\cite{BR16}. Due to lack of space we only describe their properties that are necessary for performing the delay analysis. Secret sharing codes require the division of $A$ into $k-1$ row blocks and encodes them into $n$ shares of dimension $m/(k-1)\times \ell$ each. Any $k$ shares can decode $A$. Whereas, Staircase codes require the division of $A$ into $(k-1)\alpha$ row blocks, $\alpha=\text{LCM}\{k,\dots,n-1\}$, and encodes them into $n$ shares. Each share consists of $\alpha$ subshares and is of dimension $m/(k-1)\times \ell$. Any $(k-1)/(d-1)$ fraction of any $d$ shares can decode $A$, where $d\in\{k,\dots,n\}$. We show that Staircase codes outperform classical codes in terms of incurred delays.

{\em Delay model:} Let $T_A$ be the random variable representing the time spent to compute $A\mathbf{x}$ at one worker. We assume a mother runtime distribution $F_{T_A}(t)$ that is exponential\footnote{Our analysis remains true for the shifted exponential model \cite{speeding,TOFEC}.} with rate $\lambda$. Due to the encoding, each task given to a worker is $k-1$ times smaller than $A$. Let $T_i,\ i \in\{1,\dots,n\}$ denote the time spent by worker $\w_i$ to execute its task, then we assume that $F_{T_i}$ is a scaled distribution of $F_{T_A}$, i.e.,
\begin{equation*}
F_{T_i}(t)\triangleq F_{T_A}((k-1)t)=1-e^{-(k-1)\lambda t}.
\end{equation*}
For an $(n,k)$ system using Staircase codes, we assume that $T_i$ is evenly distributed between the subshares, i.e., the time spent by a worker $\w_i$ on one subshare is equal to $T_i/\alpha$. Let $T_{(i)}$ be the $i^{th}$ order statistic of the $T_i$'s and $T_{\text{SC}}$ be the time the master waits until it can decode $A\mathbf{x}$. We can write
\begin{equation*}
T_{\text{SC}}=\min_{d\in\{k,\dots,n\}}\left\{\dfrac{k-1}{d-1}T_{(d)}\right\}\triangleq \min_{d\in\{k,\dots,n\}}\alpha_d T_{(d)},
\end{equation*}
where $\alpha_i\triangleq (k-1)/(i-1)$. For an $(n,k)$ system using classical secret sharing codes, we can write $T_{\text{SS}}=T_{(k)}.$

\section{Main Results} \label{sec:main}

Our main results are summarized as follows. We provide an upper bound and a lower bound on the mean waiting time of $\m$ in Theorem~\ref{thm:main1}.
\begin{theorem}\label{thm:main1}
The mean waiting time $\mathbb{E}[T_{\text{SC}}]$ of an $(n,k)$ system using Staircase codes is upper bounded by
\begin{equation}\label{eq:main1}
\E[T_{\text{SC}}] \leq \min_{d\in\{k,\dots,n\}}\left(\frac{H_n - H_{n-d}}{\lambda(d-1)}\right),
\end{equation}
where $H_n$ is the $n^{\text{th}}$ harmonic sum defined as $H_n \triangleq \sum_{i=1}^n \frac{1}{i}$, and $H_0\triangleq0$.
The mean waiting time is lower bounded by
\begin{align} \label{eq:main2}
\E[T_{\text{SC}}] &\geq \max_{d \in \{k,\dots,n\}} 
\sum_{i=0}^{k-1}\binom{n}{i}\sum_{j=0}^{i}\binom{i}{j}(-1)^j\frac{L(d,i,j)}{\lambda}, \nonumber\\
L(d,i,j)
&=\frac{2}{n(n-1)+d(d-1)-2(i-j)(d-1)}.
\end{align}
\end{theorem}

\noindent{\em Discussion:} Our extensive simulations show that \eqref{eq:main1} is a good approximation of the mean waiting time. Moreover, by taking $d=k$ in \eqref{eq:main1}, the upper bound on the mean waiting time of Staircase codes becomes the one of classical secret sharing, i.e.,\vspace{-0.1cm}\begin{equation}
\E[T_{\text{SC}}]\leq \E[T_{\text{SS}}]=\frac{H_n - H_{n-k}}{\lambda(k-1)}.
\end{equation}
While finding the exact expression of the mean waiting time for any $(n,k)$ system remains open, we derive in Corollary~\ref{corr1} an expression for systems with $1$ and $2$ parities, i.e. $(k+1,k)$ and $(k+2,k)$ systems, using the result of Theorem~\ref{thm:main2}. Using Corollary~\ref{corr1} one can compare the performance of Staircase codes an secret sharing codes. For instance, in a $(4,2)$ system Staircase codes reduce the mean waiting time by $40\%$. 
\begin{theorem}\label{thm:main2}
Let $t_i\triangleq t(i-1)/(k-1)$, the CDF of the waiting time $T_{\text{SC}}$ of an $(n,k)$ system using Staircase codes is given by
\begin{equation}
F_{T_{\text{SC}}}(t)=1-n!\int_{y\in A(t)}\frac{F(y_k)^{k-1}}{(k-1)!}dF(y_n)\cdots dF(y_{k}),
 \end{equation}
 where $A(t)=\cap_{i\geq k}\{y_i\in(t_i,y_{i+1}]\}$ and $F(y_i)=F_{T_i}(y_i)$.
\end{theorem}

\begin{figure}[h!]
\centering
 \setlength\figureheight{0.15\textwidth}
  \setlength\figurewidth{0.4\textwidth}
  \vspace{-0.2cm}
\resizebox{0.35\textwidth}{!}{
\definecolor{mycolor1}{rgb}{1.00000,0.00000,1.00000}
\begin{tikzpicture}

\begin{axis}[width=0.951\figurewidth,
height=\figureheight,
at={(0\figurewidth,0\figureheight)},
scale only axis,
xmin=2,
xmax=30,
xlabel={Number of workers $n$},
ymin=0,
ymax=0.6,
ylabel={Mean waiting time},
axis background/.style={fill=white},
title style={font=\bfseries},
legend style={legend cell align=left,align=left,draw=white!15!black}
]
\addplot [color=red,solid,mark=o,mark options={solid}]
  table[row sep=crcr]{4	0.238095238095238\\
6	0.205688429217841\\
8	0.180636776751595\\
10	0.161654656307091\\
12	0.146807706614232\\
14	0.134844284363862\\
16	0.125381932881933\\
18	0.11735929871743\\
20	0.110407350375983\\
25	0.0964982574063961\\
30	0.0860340389972549\\
};
\addlegendentry{Lower bound in \eqref{eq:main1}};

\addplot [color=green!50!black,solid,mark=+,mark options={solid}]
  table[row sep=crcr]{4	0.541666666666667\\
6	0.316666666666667\\
8	0.243571428571429\\
10	0.204138321995465\\
12	0.178134519801186\\
14	0.159232938778393\\
16	0.144671461017615\\
18	0.133007205213088\\
20	0.123396450420217\\
25	0.105270826261523\\
30	0.0924069307748293\\
};
\addlegendentry{Upper bound in \eqref{eq:main2}};

\addplot [color=black,solid,mark=square,mark options={solid}]
  table[row sep=crcr]{4	0.412698412698413\\
6	0.287089788436538\\
8	0.231849278705714\\
10	0.198016471139951\\
12	0.17444214142825\\
14	0.156796702675244\\
16	0.142961603452222\\
18	0.131751548359067\\
20	0.122441760770835\\
25	0.104733922585574\\
30	0.0920703989195758\\
};
\addlegendentry{Mean waiting time in \eqref{eq:corr2}};
\end{axis}
\end{tikzpicture}
}
\caption{Bounds on mean waiting time $\E[T_{\text{SC}}]$ for $(k+2,k)$ systems with $\lambda=1$.}
\label{fig:thm}
\vspace{-0.2cm}
\end{figure}
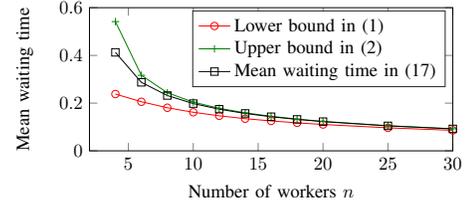

To check the tightness of the bounds we plot in Figure~\ref{fig:thm} the upper bound in~\eqref{eq:main1}, lower bound in~\eqref{eq:main2} and the exact mean waiting time in \eqref{eq:corr2} for $(k+2,k)$ systems. 

\noindent{\em Asymptotics:} To better understand the above results, we look at the asymptotic behavior of the lower and upper bounds when $n$ goes to infinity in two regimes: \begin{enumerate*}\item For a constant number of parities $r=n-k$. The mean waiting time of the system is given by
$\lim_{n \to \infty} \E[T_{SC}]=\E[T_{SS}].$
Meaning, in this regime there is no advantage in using Staircase codes (Figure~\ref{fig:thm}). %
\item For a fixed rate $R=k/n$. The mean waiting time can be bounded by $\E[T_{SC}] \leq \log\left(1/(1-c)\right)/(\lambda(nc-1))$, where $c$ is a constant satisfying $R\leq c<1$. In this regime, the mean waiting time of systems using Staircase codes is smaller by a constant factor $s$, $s<1/R$, than systems using classical secret sharing codes (Figure~\ref{fig:savings}).\end{enumerate*}

\section{Proof of Theorem~\ref{thm:main1}} \label{sec:proof1}
We will need the following characterization of order statistics for \emph{iid} exponential random variables. 
\begin{theorem}[Renyi \cite{Renyi53}] \label{thm:Renyi}The $d^{\text{th}}$ order statistic $T_{(d)}$ of $n$ iid exponential random variables $T_i$, with distribution function ${F}(t)=1-e^{-\lambda t}$, %
is equal to a random variable $Z$ in the distribution, where \vspace{-0.3cm}%
\begin{align*}
T_{(d)} &\triangleq \sum_{j=0}^{d-1}\frac{Z_j}{n-j},
\vspace{-0.3cm}
\end{align*}
and $Z_j$ are iid random variables with distribution ${F}(t)$.
\end{theorem}

\subsection{Upper bound on the mean waiting time}
We use Jensen's inequality to upper bound the mean waiting time $\E[T_{\text{SC}}]$. The exact mean waiting time is given by
\begin{equation*}
\E[T_{\text{SC}}]=\E\left[\min_{d\in\{k,\dots,n\}} \left\{\frac{k-1}{d-1}T_{(d)}\right\}\right].
\end{equation*}
Since $\min$ is a convex function, we can use Jensen's inequality to write%
\begin{equation}\label{eq:jensen}
\E\left[\min_{d\in\{k,\dots,n\}} \left\{\frac{k-1}{d-1}T_{(d)}\right\}\right]\leq \min_{d\in\{k,\dots,n\}}\left\{\E\left[ \frac{k-1}{d-1}T_{(d)}\right]\right\}.
\end{equation}
The average of the $d^{\text{th}}$ order statistic $\E[T_{(d)}]$ can be written as
\begin{align}
\E[T_{(d)}]&=\E[Z_i]\sum_{j=0}^{d-1}\dfrac{1}{n-j}
=\dfrac{H_n-H_{n-d}}{\lambda(k-1)}\label{eq:avg}
\end{align}
Equations~\eqref{eq:jensen} and~\eqref{eq:avg} conclude the proof. We give an intuitive behavior of the upper bound. The harmonic number can be approximated by $H_n\approx \log(n)+\gamma,$ where $\gamma \approx 0.577218$ is called the Euler-Mascheroni constant. Therefore, $\log(n)<H_n< \log(n+1)$. Hence, we can write
\begin{align}\label{eq:finavg}
\mathbb{E}[T_{\text{SC}}]<\min \left\{
\begin{aligned}
&\min_{d\in\{k,\dots,n-1\}}\left\{\dfrac{1}{\lambda(d-1)}\log\left(\dfrac{n+1}{n-d}\right)\right\},\\
&\dfrac{1}{\lambda(n-1)}\log\left(n+1\right)
\end{aligned}\right\}.
\end{align}

\subsection{Lower bound on the mean waiting time}
To lower bound the mean waiting time $\E[T_{\text{SC}}]$, we find the probability distribution of a small (sufficient) set of conditions that result in $T_{\text{SC}}>t$. This distribution serves as a lower bound on the exact distribution of $T_{\text{SC}}$. For a given $d\in \{k,\dots,n\}$, consider the following set of conditions\begin{align*}
\mathcal{C}\triangleq \left\{T_{(k)} > \frac{t}{\alpha_d}\right\}\bigcap_{j=d+1}^{n}\left\{T_{(j)}-T_{(j-1)} > \frac{t}{\alpha_j} - \frac{t}{\alpha_{j-1}}\right\},
\end{align*}
where $\alpha_j\triangleq (k-1)/(j-1)$. For $T_{\text{SC}}$ to be greater than $t$, all the $j^{\text{th}}$ order statistic $T_{(j)}$'s must be greater than $t/\alpha_j$ for $j\in\{k,\dots,n\}$. We show that if $\mathcal{C}$ is satisfied, then the previous condition is satisfied. If $T_{(k)}>t/\alpha_d$, then $T_{(i)}>t/\alpha_i$ for all $i\in\{k,\dots,d\}$, because $T_{(i)}\geq T_{(k)}>t/\alpha_d>t/\alpha_{i}$. It follows that if for all $j\in\{d+1,\dots,n\},$ $T_{(j)}-T_{(j-1)}>t/\alpha_j-t/\alpha_{j-1}$, then $T_{(j)}>t/\alpha_j$. Therefore, $\Pr \left(T_{\text{SC}}>t\right) \geq \Pr\left(\mathcal{C} \text{ is satisfied}\right)\triangleq \Pr(\mathcal{C})$. Furthermore,
\begin{equation}\label{eq:ub1}
\E[T_{\text{SC}}]=\int_{0}^{\infty}\Pr \left(T_{\text{SC}}>t\right)dt\geq \int_{0}^{\infty} \Pr(\mathcal{C})dt.
\end{equation} Next we derive an expression of $\int_{0}^{\infty}\Pr\left(\mathcal{C} \right)dt$. Note that $1/\alpha_j-1/\alpha_{j-1}=1/(k-1)$, using Theorem~\ref{thm:Renyi} we can write
\begin{equation}\label{eq:ub2}
\Pr\left\{T_{(j)}-T_{(j-1)}>\frac{t}{k-1}\right\}=\bar{F}_{Z_j}\left(\frac{(n-j+1)t}{k-1}\right),
\end{equation}
where $\bar{F}_{Z_j}(t)\triangleq \Pr\left(Z_j>t\right)$. From~\eqref{eq:ub2} we get
\begin{equation}\label{eq:ub3}
\Pr\left(\mathcal{C} \right)= \bar{F}_{T_{(k)}}\left(\frac{t}{\alpha_d}\right)\prod_{j=d+1}^{n}\bar{F}_{Z_j}\left(\frac{(n-j+1)}{(k-1)}t\right)
\end{equation}
\noindent Since $\bar{F}_{Z_j}(t)= e^{-(k-1)\lambda t}$, we can write
\begin{align}
\prod_{j=d+1}^{n}\bar{F}_{Z_j}\left(\frac{(n-j+1)}{(k-1)}t\right)&=\bar{F}_{Z_j}\left(\sum_{j=d+1}^{n}\frac{(n-j+1)}{(k-1)}t\right) \nonumber\\
&=\bar{F}_{Z_j}\left(t \frac{(n-d)(n-d+1)}{2(k-1)}\right)\label{eq:ub4}.
\end{align}
On the other hand, $\bar{F}_{T_{(k)}}\left(t/\alpha_d\right)$ is the probability that there are at most $k-1$ $T_i$'s less than $t/\alpha_d$, therefore
\begin{align}\label{eq:ub5}
\bar{F}_{T_{(k)}}\left(\frac{t}{\alpha_d}\right) &=\sum_{i=0}^{k-1}\binom{n}{i}F_{T_i}\left(\frac{t}{\alpha_d}\right)^i \bar{F}_{T_i}\left(\frac{t}{\alpha_d}\right)^{n-i}.
\end{align}
Recall that ${F}_{T_i}(t)=1-e^{(k-1)\lambda t}=1-\bar{F}_{T_i}(t)$, therefore by using the binomial expansion we can write
\begin{align}\label{eq:ub6}
F_{T_i}\left(\frac{t}{\alpha_d}\right)^i=\sum_{j=0}^{i}\binom{i}{j}(-1)^{j}\bar{F}_{T_i}\left(\frac{t}{\alpha_d}\right)^{j}.
\end{align}
Using~\eqref{eq:ub6} and the fact that $F_{T_i}(t)=e^{-(k-1)\lambda t}$, \eqref{eq:ub5} becomes\begin{align}\label{eq:ub7}
F_{T_{(k)}}\left(\frac{t}{\alpha_d}\right)=\sum_{i=0}^{k-1}\binom{n}{i}\sum_{j=0}^{i}\binom{i}{j}(-1)^{j}\bar{F}_{T_i}\left(t\frac{(n-i+j)(d-1)}{(k-1)}\right).
\end{align}
Combining~\eqref{eq:ub4} and \eqref{eq:ub7} and noting that $\bar{F}_{T_i}(t)=\bar{F}_{Z_j}(t)=e^{-\lambda(k-1)t}$, \eqref{eq:ub3} becomes
\begin{align}\label{eq:ub8}
\Pr\left(\mathcal{C}\right)&= \sum_{i=0}^{k-1}\binom{n}{i}\sum_{j=0}^{i}\binom{i}{j}(-1)^{j}\exp\big(-\lambda t(n-i+j)(d-1) \nonumber \\
& ~ - \lambda t (n-d)(n-d+1)/2\big).
\end{align}
Note that $\int_{0}^{\infty}e^{-xt}dt=1/x$ and that the integral of a sum is equal to the sum of the integrals. Therefore, integrating \eqref{eq:ub8} from $0$ to $\infty$ and maximizing it over all values of $d,\ d\in\{k,\dots,n\}$, concludes the proof.

\begin{figure*}[!t]
\normalsize
\setcounter{MYtempeqncnt}{\value{equation}}
\setcounter{equation}{15}
\begin{equation}
\label{eq:corr1}
\mathbb{E}\left[T_{\text{SC}}\right]
= \frac{1}{\lambda}\sum_{i=2}^{k+1}(-1)^{i}\binom{k+1}{i}\left[\frac{i}{k+(k-1)(i-1)}-\dfrac{1}{k i} \right].
\vspace{-0.4cm}
\end{equation}
\begin{align}
\label{eq:corr2}
\E[T_{\text{SC}}]&=\sum_{i=2}^{k+2}\frac{(-1)^i\binom{k+2}{i}}{\lambda}\left[\dfrac{i}{(k+1)+ k(i-1)}-\dfrac{1}{(k+1)i}+\dfrac{i(i-1)}{4(k+1)+2(k-1)(i-2)}-\dfrac{i(i-1)}{(2k+1)+(k-1)(i-2)}\right].
\end{align}
 \setcounter{equation}{\value{MYtempeqncnt}}
\hrulefill
\vspace{-0.5cm}
\end{figure*}
\section{Proof of Theorem~\ref{thm:main2}}   \label{sec:proof2}
We derive an integral expression leading to the probability distribution of the waiting time $T_{\text{SC}}$. Since the delays at the workers' side $T_i$'s are independent and are absolutely continuous with respect to the Lebesgue measure 
(i.e. the probability density exists), we have
\begin{align*}
f_{T_{(1)}, \ldots, T_{(n)}}(t_1,\ldots, t_n) = n! \prod_{i=1}^nf_{T_i}(t_i) = n!\lambda^n \exp\left(-\lambda \sum_{i=1}^{n}t_i\right),
\end{align*}
where $t_i$ denotes $t/\alpha_i$ and $0 \leq t_1 \leq \ldots \leq t_n$.
Therefore we can write the distribution of $T_{\text{SC}}$ as%
\begin{align*}
\Pr\{T_{\text{SC}} > t\} &= \Pr\bigcap_{d=k}^{n}\left\{T_{(d)} > t_d\right\}= \int_{A(t)}f_{T_{(1)},\dots,T_{(n)}}(y) dy,
\end{align*}
where $y_{n+1} = \infty$ and
\begin{align*}
A(t)&=\left\{0 \leq  y_1 \leq \ldots \leq y_n: y_d > t_d, \text{ for }k \leq d \leq n\right\}\\
&=\cap_{i \geq k}\{y_i \in (t_i, y_{i+1}]\}\cap_{i < k}\{y_i \in [0, y_{i+1}]\}.
\end{align*}
That is, we can re-write $\Pr\{T_{\text{SC}}>t\}$ as
\begin{align*}
&n!\int_{t_n}^{\infty}\cdots\int_{t_k}^{y_{k+1}}\prod_{i=k}^ndF_{T_i}(y_i)\left(\int_{0}^{y_k}\cdots\int_{0}^{y_{2}}\prod_{i=1}^{k-1}dF_{T_i}(y_i)\right). %
\end{align*} 

\begin{claim}\label{claim:int}
$\int_{0}^{y_k}\cdots\int_{0}^{y_{2}}\prod_{i=1}^{k-1}dF_{T_i}(y_i) = \frac{F(y_k)^{k-1}}{(k-1)!}.$
\end{claim}

\noindent The result of Claim~\ref{claim:int} is straightforward, it follows from integrating $k-1$ times the complementary CDF of an exponential random variable in respect to its derivative. This completes the proof. A more detailed proof of Claim~\ref{claim:int} can be found in \cite{BPR17}. We state the mean waiting time for the $(k+2,k)$ and $(k+1,k)$ systems in Corollary~\ref{corr1}.
\begin{corollary}\label{corr1}
The mean waiting time $\E[T_{\text{SC}}]$ for $(k+1,k)$ and $(k+2,k)$ systems are given by \eqref{eq:corr1} and \eqref{eq:corr2}, respectively.
\end{corollary}
\vspace{-0.2cm}

\vspace{-0.1cm}
\section{Simulations} \label{sec:simu}
We check the tightness of the bounds of Theorem~\ref{thm:main1} and measure the improvement, in terms of delays, of Staircase codes over classical secret sharing codes for systems with fixed rate $R\triangleq k/n$. In Figure~\ref{fig:sims} (a) we plot the upper bound~\eqref{eq:main1}, lower bound~\eqref{eq:main2} and the simulated mean waiting time for $R=1/4$. Our extensive simulations show that the upper bound is a good approximation of the exact mean waiting time, whereas the lower bound might be loose.
\begin{figure}[h!]
\vspace{-0.3cm}
\centering
    \setlength\figureheight{0.3\textwidth}
  \setlength\figurewidth{0.35\textwidth}
  \begin{minipage}[b]{0.22\textwidth}
  \centering
  \captionsetup{subtype,width=0.95\textwidth}
\resizebox{1.1\textwidth}{!}{\definecolor{mycolor1}{rgb}{1.00000,0.00000,1.00000}
\begin{tikzpicture}

\begin{axis}[width=0.951\figurewidth,
height=\figureheight,
at={(0\figurewidth,0\figureheight)},
scale only axis,
xmin=0,
xmax=100,
xlabel={Number of workers $n$},
ymin=0,
ymax=0.3,
ylabel={Mean waiting time},
ylabel style={at={(0.08,0.5)}},
axis background/.style={fill=white},
title style={font=\bfseries},
legend style={legend cell align=left,align=left,draw=white!15!black}
]
\addplot [color=red,solid,mark=o,mark options={solid}]
  table[row sep=crcr]{8	0.0521821631878558\\
12	0.0307960381511372\\
24	0.0136747350837323\\
40	0.00779753422894387\\
80	0.00374860403851571\\
100	0.00297454836203638\\
120	0.00246524385016883\\
180	0.00162842143895292\\
200	0.00146286367323347\\
};
\addlegendentry{Lower bound in~\eqref{eq:main1}};

\addplot [color=green!50!black,solid,mark=+,mark options={solid}]
  table[row sep=crcr]{8	0.211507936507937\\
12	0.127588383838384\\
24	0.0560676091854582\\
40	0.0315062120017761\\
80	0.0150320455786199\\
100	0.0119175786286072\\
120	0.00987233394844413\\
180	0.00651723061286255\\
200	0.0058540889689299\\
};
\addlegendentry{Upper bound in~\eqref{eq:main2}};

\addplot [color=black,solid]
  table[row sep=crcr]{8	0.152152771438816\\
12	0.0959340531352646\\
24	0.046358775327669\\
40	0.0276826602386197\\
80	0.0139559984547979\\
100	0.011189349164409\\
120	0.00935173993145852\\
180	0.00626077504499385\\
200	0.00564120834589456\\
};
\addlegendentry{Staircase codes};

\addplot [color=mycolor1,solid,mark=asterisk,mark options={solid}]
  table[row sep=crcr]{8	0.267293482980229\\
12	0.137140346920711\\
24	0.0561376894836278\\
40	0.0314633684805783\\
80	0.0150542651311887\\
100	0.0119258213000058\\
120	0.00988636681494367\\
180	0.00652176891712045\\
200	0.00585718230089935\\
};
\addlegendentry{Secret sharing};

\end{axis}
\end{tikzpicture}}
\caption{Waiting time for systems with rate $R=1/4$.}
\label{fig:sim1}
\end{minipage}\hfill\begin{minipage}[b]{0.235\textwidth}
\centering
\resizebox{\textwidth}{!}{\definecolor{mycolor1}{rgb}{1.00000,0.00000,1.00000}
\begin{tikzpicture}

\begin{axis}[width=0.951\figurewidth,
height=\figureheight,
at={(0\figurewidth,0\figureheight)},
scale only axis,
xmin=2,
xmax=30,
xlabel={Number of workers $n$},
ymin=0,
ymax=0.6,
ylabel={Mean waiting time},
ylabel style={at={(0.08,0.5)}},
axis background/.style={fill=white},
title style={font=\bfseries},
legend style={legend cell align=left,align=left,draw=white!15!black}
]
\addplot [color=red,solid,mark=o,mark options={solid}]
  table[row sep=crcr]{4	0.238095238095238\\
6	0.205688429217841\\
8	0.180636776751595\\
10	0.161654656307091\\
12	0.146807706614232\\
14	0.134844284363862\\
16	0.125381932881933\\
18	0.11735929871743\\
20	0.110407350375983\\
25	0.0964982574063961\\
30	0.0860340389972549\\
};
\addlegendentry{Lower bound in \eqref{eq:main1}};

\addplot [color=green!50!black,solid,mark=+,mark options={solid}]
  table[row sep=crcr]{4	0.541666666666667\\
6	0.316666666666667\\
8	0.243571428571429\\
10	0.204138321995465\\
12	0.178134519801186\\
14	0.159232938778393\\
16	0.144671461017615\\
18	0.133007205213088\\
20	0.123396450420217\\
25	0.105270826261523\\
30	0.0924069307748293\\
};
\addlegendentry{Upper bound in \eqref{eq:main2}};

\addplot [color=black,solid,mark=square,mark options={solid}]
  table[row sep=crcr]{4	0.412698412698413\\
6	0.287089788436538\\
8	0.231849278705714\\
10	0.198016471139951\\
12	0.17444214142825\\
14	0.156796702675244\\
16	0.142961603452222\\
18	0.131751548359067\\
20	0.122441760770835\\
25	0.104733922585574\\
30	0.0920703989195758\\
};
\addlegendentry{Mean waiting time in \eqref{eq:corr2}};

\end{axis}
\end{tikzpicture}}
\vspace{-0.45cm}
\captionsetup{subtype,width=0.95\textwidth}
\caption{Delay reduction using Staircase codes.}
\label{fig:savings}
\end{minipage}
\caption{Simulations for $(n,k)$ systems with fixed rate.}\label{fig:sims}
\vspace{-0.2cm}
\end{figure}

\noindent Figure~\ref{fig:sims} (b) aims to better understand the comparison between Staircase codes and classical codes. We plot the normalized difference between the mean waiting times, i.e., $\left(\E[T_{\text{SS}}]-\E[T_{\text{SC}}]\right)/\left(\E[T_{\text{SS}}]\right)$, for different rates. For high rates, Staircase codes offer high savings for small values of $n$, whereas for low rates Staircase codes offer high savings for all values of $n$.

\bibliographystyle{ieeetr}
\bibliography{Rawad}

\end{document}